*Review*

# Edge Computing for IoT-Enabled Smart Cities: A Review of Middleware, Explainable AI, and User Behavior Analytics

**Alexandros Gazis [1,2], George Vavouras [3], Nikos E. Mastorakis [4], Kleanthi Santamouri[5], Nikolaos Pournaras[6] and Stylianos Pappas [7,*]**

[1] Edinburgh Business School, Heriot-Watt University, Edinburgh EH14 4AS, UK
[2] Department of Electrical and Computer Engineering, School of Engineering, Democritus University of Thrace, 67100 Xanthi, Greece
[3] Department of Electrical and Computer Engineering, School of Engineering, Aristotle University of Thessaloniki, 54124 Thessaloniki, Greece
[4] English Language Faculty of Engineering, Technical University of Sofia, 1756 Sofia, Bulgaria
[5] English Language Teacher, Private Primary Education, Athens, Greece
[6] Salvezza Energy Systems LTD, Sirakouson 4, Paphos 8016, Cyprus
[7] Department of Electrical Engineering and Computer Science, Hellenic Naval Academy, Terma Chatzikyriakou, 18539 Piraeus, Greece
* steliospappas@teemail.gr

**Abstract**

This review looks at cloud–IoT edge middleware for analyzing user behavior in smart cities, with a particular focus on linking behavioral analytics to energy-aware and mission-critical urban services. The topic sits at the crossroads of smart-city platforms, IoT sensing, edge/fog/cloud computing, service-oriented architectures, machine learning, explainable AI, privacy-aware governance, and energy-critical applications such as EV charging, energy demand, environmental exposure, crowd safety, and the operational resilience of urban infrastructure. As such, this paper maps out the main categories of user-behavior data, examines five application domains, compares several case studies, and puts forward a new XAI-EdgeSOA Middleware Framework called X-EdgeCity Middleware. This framework brings together IoT sensing, edge preprocessing, fog coordination, event-driven communication, cloud analytics, XAI explanations, privacy-by-design, and benchmark logging. The goal is to support not just general smart-city applications, but also energy-sensitive or mission-critical services, where latency, reliability, explainability, and accountability have a direct impact on urban safety and continuity of operations.

## 1. Introduction

The smart city has moved on from its early, tech-heavy model of "enhanced" urban infrastructure into a much more complex socio-technical ecosystem, one where sensors, mobile devices, cloud infrastructure, edge/fog nodes, middleware platforms, data analytics services, and citizen-facing apps all connect into ongoing flows of information and decision-making. Within this setting, analyzing user behavior - residents, visitors, workers, and people just passing through - isn't a side function of urban management anymore. It's central to turning the city into an adaptive, predictive, participatory environment. This matters even more once behavior gets tied to energy-aware and mission-critical services: things like EV charging, energy demand management, environmental exposure, crowd safety, and keeping critical urban infrastructure running.

This transition is made possible through the implementation of a new emergent technology composed of IoT, cloud, edge/fog computing, and middleware architecture. The InterSCity project defines a smart city as an environment supported by ICT in social, operational, and technological terms, enhancing the citizen experience and providing integrated, sustainable, and affordable services [1]. According to the literature, existing smart city implementations are based on proprietary systems that lack adequate interoperability, portability, extensibility, and cost-effectiveness. Thus, reusable middleware platforms are needed [1,2].

The middleware analyses the data captured by the sensors and allows extraction of value-adding information. It deals with ingestion of data, composition of services, stream processing, storage, interoperability, API exposure, context awareness, security, privacy controls, and runtime adaptation. In this model, the various connections between physical space, sensors, software services, and analytical models are analyzed, and user behavior is studied as an emerging consequence. Specifically, it includes repeated movement patterns and route choices, the duration of stay at a point of interest, use of public infrastructure, interaction with recreational space, interaction with environmental conditions, and connection with digital or augmented interfaces [2-5].

The practical uses vary greatly from one another. In transportation, the waiting time at stops and hubs can help in route planning. The spatial distribution of users within such spaces and places plays an important role in the design of settings [5]. Records of EV charging behavior, charging station usage and parking patterns, as well as user adaptability to environmental stress, count as fundamental data for smart-city services. For high-stakes endeavors, such as crowd management, emergency response, and building resiliency, these sorts of analyses have to meet much stricter requirements in terms of latency, reliability, explainability, and governance since algorithmic output here can directly affect urban safety and continuity of operations [6].

According to the literature, machine learning techniques have been linked to resident-behavior analysis. An IEEE study attributes the organization of resident behavior data to various resident behavior dimensions like travel mode, consumption, and social behavior. It uses SVM, Random Forest, and LSTM for classification and prediction [4]. Just because we can collect the data, it doesn’t mean we should. The streams of data that we deal with in smart cities are usually very large, heterogeneous, incomplete, and noisy. For this reason, middleware must include features such as preprocessing, missing-data imputation, temporal aggregation, spatial analysis, and visualization [6].

Understanding user behavior raises major data-privacy, transparency, and accountability concerns. A smart city can easily transform from a service-improving environment to a surveillance one, especially if it begins to collect images, location traces, presence data, or inference data around people’s preferences and intentions. Citizens are very concerned about privacy and ethics when their personal data is collected in smart-city data collection scenarios. The type of data we collect, who has access to it, and how long it is retained shape whether these systems will be accepted or rejected [7,8]. Examining the cloud-IoT middleware and user behavior cannot be exclusively a technical matter. It should incorporate mechanisms to ensure data minimization, explainability, privacy by design, consent management, and participatory governance.

The study is guided by the following research questions:

- RQ1. Which cloud–IoT, edge/fog, and middleware architectures are used to collect, process, and analyze user behavior in smart cities?
- RQ2. Which evaluation metrics do the case studies use to assess solution performance, both at the cloud/hardware/request level and at the behavioral-analytics level?
- RQ3. How can a new middleware framework be designed that combines IoT edge computing, SOA/microservices, XAI-driven analytics, and privacy-aware governance for smart-city applications?

Our research landscape is summarized in Table 1. Specifically, this table presents the main areas covered in this review, including infrastructure, middleware, behavioural analytics, adaptive interfaces, and governance. For each area, it summarises the main focus and the key challenge addressed in the study.

**Table 1.** Scope of the review

| Dimension | Focus | Main challenges |
|---|---|---|
| Infrastructure | Cloud, edge, fog, IoT networks, and sensors | Latency, availability, and scalability |
| Middleware | Data streams, APIs, service composition, and interoperability | Bringing heterogeneous data sources together |
| Behavioural analytics | Proximity, dwell time, mobility, and point-of-interest interaction | Turning raw data into useful knowledge |
| Adaptive interfaces | Mobile apps, augmented reality, digital twins, and visualisation tools | Providing feedback to users and city services |
| Governance | Privacy, transparency, consent, and participation | Avoiding surveillance and unfair algorithmic decisions |

## 2. Review Methodology & PRISMA framework statement

The present research work is conducted as a scoping review based on the PRISMA system. The research topic's nature influences the research methodology. The domain of cloud–IoT edge middleware that analyzes user behavior in smart cities cannot be described as a single, closed, and mature area of research. It lies at the intersection of quite a few much more established sub-fields: smart-city platforms, IoT middleware, cloud/edge/fog computing, user behaviour analysis, spatial sensing, machine learning, explainable AI, privacy and participatory governance.

The reporting of the search and selection process follows the general principles of PRISMA 2020 [9] and the mapping nature of the review is reflected in PRISMA-ScR [10]. The principles of "scoping-study" [11,12] are also factored in. This strategy is effective in disciplines where there is inconsistency in the use of terminology - in any event, where the same technical practice has other names.

The search process drew on several main academic databases and publishing platforms: IEEE Xplore, SpringerLink, MDPI, Taylor & Francis Online, ACM Digital Library, and ScienceDirect/Elsevier. Indexes and publisher pages such as Wiley Online Library, PLOS, SAGE Journals, Nature, BMJ, VLDB, AAAI/AI Magazine, and official journal pages were also selectively checked. Sources without a DOI were excluded from the final bibliography.

In the first stage of the search, 25 initial top keywords were drawn up, based on the title, the research questions, the application domains, and core bibliographic terminology. These keywords weren't just mechanical search terms - they set the initial thematic perimeter of the corpus. The 25 initial keywords are shown in Table 2.

**Table 2.** Initial keywords and corresponding thematic axes used for the literature search

| No. | Initial Keyword | Thematic Axis |
|---|---|---|
| 1. | smart city | Smart-city / urban context |
| 2. | smart cities | Smart-city / urban context |
| 3. | urban computing | Smart-city / urban context |
| 4. | future internet | Smart-city / urban context |
| 5. | cloud computing | Cloud–IoT / edge infrastructure |
| 6. | edge computing | Cloud–IoT / edge infrastructure |
| 7. | fog computing | Cloud–IoT / edge infrastructure |
| 8. | IoT sensors | IoT sensing / urban sensing |
| 9. | urban sensing | IoT sensing / urban sensing |
| 10. | IoT middleware | Middleware/platform |
| 11. | resident behaviour | User/resident behaviour |
| 12. | human behaviour | User/resident behaviour |
| 13. | citizen behaviour | User/resident behaviour |
| 14. | behavioural analytics | Behavioural analytics |
| 15. | pattern recognition | Behavioural analytics |
| 16. | dwell time | Spatial interaction |
| 17. | proximity | Spatial interaction |
| 18. | trajectory | Spatial interaction |
| 19. | PoIs | Spatial interaction |
| 20. | digital twin | Visualization / digital twin / AR |
| 21. | explainable AI | XAI / interpretability |
| 22. | XAI | XAI / interpretability |
| 23. | algorithmic transparency | Governance/privacy |
| 24. | privacy | Governance/privacy |
| 25. | participatory governance | Governance/privacy |

In the second stage, the 25 keywords were consolidated into 10 final keyword clusters, used as inclusion-oriented thematic criteria. This consolidation was necessary for three reasons: first, to avoid the search spreading too thin across generic smart-city papers; second, to tie each final source to technical infrastructure, behavioral analysis, or governance; and third, to make it feasible to systematically check titles, abstracts, and keywords. The 10 inclusion clusters and their rationale are shown in Table 3.

**Table 3.** Final keyword clusters used as inclusion criteria and their rationale

| No. | Keyword Cluster | Final Terms / Concepts | Rationale for Inclusion |
|---|---|---|---|
| 1. | Smart-city / urban context | smart city, smart cities, urban computing, future city | Ensures that the selected article focuses on an urban setting rather than a generic IoT, cloud, or computing environment. |
| 2. | User/resident behaviour | user behaviour, resident behaviour, human behaviour, citizen behaviour | Links the selected source to the core topic of behavioural analysis in urban environments. |
| 3. | IoT sensing / urban sensing | Internet of Things, IoT sensors, urban sensing, smart objects | Checks whether the study includes a mechanism for collecting data from physical space, connected devices, or urban sensor systems. |
| 4. | Middleware/platform | middleware, platform, SOA, microservices, service composition | Identifies sources related to the intermediary software layer that connects sensing, processing, and application services. |
| 5. | Cloud/edge/fog | cloud computing, edge computing, fog computing, mobile cloud | Ensures relevance to the computing architecture used for storage, processing, and real-time urban analytics. |
| 6. | Behavioural analytics | machine learning, deep learning, pattern recognition, predictive analytics | Filters sources that include an analytical method, behavioural model, or data-driven prediction mechanism. |
| 7. | Spatial interaction | POI, proximity, dwell time, trajectory, pedestrian flow | Connects behavioural data with spatial and temporal features of movement, presence, and interaction. |
| 8. | Visualization / digital twin / AR | dashboard, visualization, digital twin, augmented reality, AR | Identifies papers that return processed information to users, planners, or decision-makers through visual or immersive interfaces. |
| 9. | XAI / interpretability | explainable AI, XAI, interpretability, transparent analytics | Covers the explainability, transparency, and model-auditing dimension of AI-based urban analytics. |
| 10. | Governance/privacy | Privacy, ethics, consent, algorithmic transparency, participatory governance | Ensures that the corpus includes issues of legitimate data use, rights, accountability, and citizen participation. |

Applying the PRISMA framework involved four distinct stages. In the identification stage, records from the selected databases were logged and duplicates removed. In the screening stage, titles, abstracts, and keywords were checked against the inclusion clusters in Table 3. In the eligibility stage, full texts were assessed, with particular focus on the relationship between architecture, data, application domain, and evaluation metrics. In the inclusion stage, only sources that could contribute to the chapters, the comparative tables, and the proposed framework were retained.

Inclusion and exclusion criteria were applied hierarchically, so the final corpus would be thematically coherent rather than simply broad. These criteria are shown in Table 4.

**Table 4.** Inclusion/exclusion criteria for the PRISMA-based screening of the review

| Screening Dimension | Inclusion Criteria | Exclusion Criteria |
| --- | --- | --- |
| Bibliographic status | Article, conference paper, review, or book chapter with a DOI and complete bibliographic data. | Sources without a DOI, blogs, slide decks, commercial white papers, or unverifiable texts. |
| Smart-city scope | Clear connection to smart cities, urban computing, smart urbanism, or public urban infrastructure. | Generic IoT or cloud-computing papers with no urban or public-spatial context. |
| Infrastructure/middleware relevance | Connection to IoT, cloud, edge/fog computing, middleware, SOA, microservices, stream processing, or smart-city platforms. | Purely sociological or urban-planning papers with no technical relevance to the topic. |
| Behavioural relevance | Analysis, prediction, or interpretation of human, user, resident, or citizen behaviour, or technical infrastructure that supports such analysis. | Plain descriptions of sensors or hardware with no link to user behaviour or urban service use. |
| Application domain | Connection to mobility, public spaces, energy/environment, public safety, tourism, or public services. | Applications outside the urban setting, or studies with no clear connection to the research questions. |
| Evaluation/benchmarking | Reference to metrics, performance evaluation, architecture, case study, prototype, testbed, or empirical assessment. | Texts with no evaluable elements, metrics, prototype, architecture, or clear methodological contribution. |
| Ethics/governance | For governance-oriented papers: connection to data collection, privacy, AI, algorithmic accountability, consent, or participation. | General regulatory discussions with no clear link to smart-city data, sensing, AI, or urban analytics. |

The selection process, and the relationship between identification, screening, eligibility, and inclusion, is shown in Figure 1.

Through this process, source selection works as a controlled field mapping: technical sources get tied to architecture and metrics, behavioral sources to data and applications, and governance sources to the conditions for legitimate data use. The synthesis of results is narrative and comparative. No meta-analysis is attempted, since the studies are heterogeneous in terms of data, architectures, metrics, and application domains. Instead, thematic synthesis is carried out based on middleware architectures, data types, user-behavior-analysis applications, AI/XAI methods, benchmarks, and privacy/governance mechanisms.

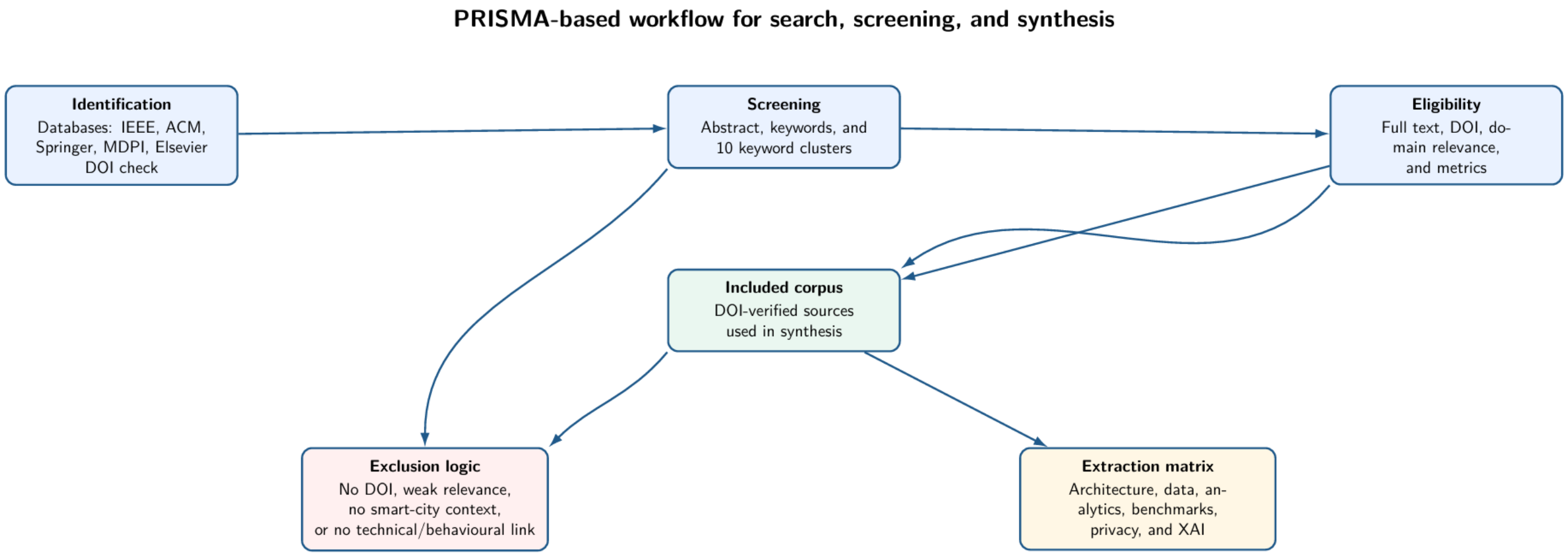

**Figure 1.** PRISMA-based process for searching, screening, and synthesizing literature

## 3. Conceptual Background: Smart Cities, Cloud-IoT Middleware and Future Internet

The need for converting urban information into useful knowledge is at the centre of the concept of a "smart city". Specifically, IoT is the basic technological foundation for data collection in cities. In a smart-city context, IoT links devices such as sensors, actuators, cameras, smart objects, mobile devices, parking systems, energy meters, environmental stations, and digital public services. Having numerous sensors alone does not constitute having a smart city. A middleware layer to structure heterogeneous sources, protocols, data formats, and applications is what is needed [15].

The integration of cloud and Internet of Things (IoT) technologies connotes the first big solution to the scaling issue. The capabilities of cloud computing include elastic storage, high-performance computing, batch analytics, trained modelling, long-term data management, and the sharing of datasets between applications [16]. Even though these benefits are significant, depending only on centralized infrastructure is not adequate for applications requiring low latency, localized processing, and immediate response. Fog and edge computing can act as interesting solutions to these [17-20].

Moreover, awareness of context is key. Sensor data points need to be analyzed and processed; otherwise, they are not useful in raw format, i.e., as provided by the output of common sensory devices and actuators. Being present at a point of interest can signal tourist activity or congestion, a wait for a public service, or a crowding risk, depending on the place, time, season, and type of activity. For IoT systems, context-aware computing is paramount because collecting, modeling, reasoning about, and disseminating context ultimately turns raw data into information [21].

Service-oriented architecture and microservices approaches matter a lot for smart-city middleware. Service-oriented computing describes services as reusable software units that can be composed to support complex processes [22,23]. Microservices take this further with smaller, independent services that can be deployed, scaled, and updated on their own [23,24]. For early prototypes or startup development, this means an initial monolithic application can launch quickly and gradually be broken down into services.

The Future Internet dimension brings in the question of scale. InterSCity organizes smart-city challenges into infrastructure, middleware, and services/applications, with an emphasis on networking, high-performance distributed computing, software engineering, and mathematical modeling [1,6]. The SmartSantander testbed is a good example of large-scale IoT experimental infrastructure deployed in a real city, built for large-scale experimentation and evaluating IoT concepts under real-life conditions [25].

The classification of middleware for the cloud, the IoT, and the edge is illustrated in Figure 2. Lastly, the conceptual layers of the architecture are summarized in Table 5.

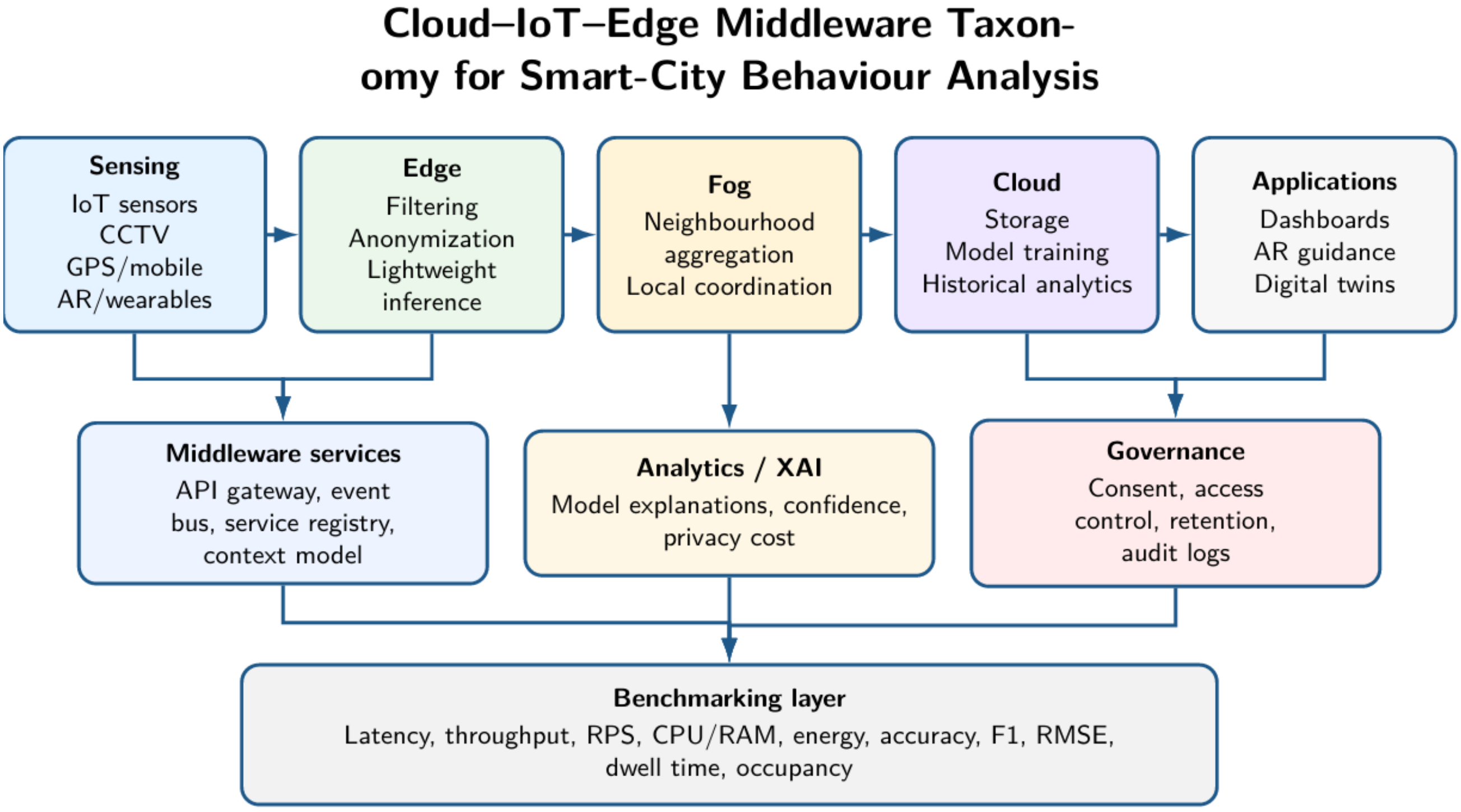

**Figure 2.** Taxonomy of cloud-IoT-edge middleware for smart-city user behavior analysis.

## 4. User Behavior Data in Smart Cities

User behavior analysis in smart cities relies on turning heterogeneous urban data into interpretable spatiotemporal patterns. Traditional urban research methods mostly relied on questionnaires, interviews, censuses, and on-site observation. Modern smart-city systems, by contrast, generate continuous, multimodal, and often real-time data. The information represents more than just the infrastructure; it also gives an indication of the way people engage with the infrastructure. For instance, how people move, where they stop, how long they stay, which POIs they visit, which services they opt for, and what is the response to the environment or traffic.

Gathering user behavior data should not be equated with recording their physical presence. A presence of a person at a certain moment in the city is a fact. As soon as it obtains a temporal value or becomes linked to a behaviour through its duration, repetition, route, type of space or service, event or environmental context, it becomes a behaviour data point. The literature on urban computing has described the city as a space where data mining, ubiquitous computing, sensing, human mobility and urban planning come together into a single field of research [26].

Mobile phones and GPS-enabled devices are among the most significant sources of spatiotemporal data. The analysis of paths, speed, stopping points, ordering patterns, points of interest, and recurrent mobility patterns is possible through GPS trajectories [27].

Social media with geotags and location-based social networks offer a new dimension by linking space, time, activity, and sometimes even self-representation [28-34]. Smart-card data for public transport can be a good source of analysis for commuting patterns, passenger clustering, and station characterization [29-37].

The traces that we get through Bluetooth and Wi-Fi don't always give the full route, but they reveal presence, repeat appearance, dwell time, pedestrian flow, and moves between sensors [38]. Lastly, it is important to note that systems of mobile crowdsensing and participatory sensing incorporate a participatory feature whereby the user serves as a possible contributing node of information [32-35].

Environmental, energy, and infrastructure data do not have behavioral significance unless linked with human presence or human use. Gadgets like air quality sensors, traffic counters, noise sensors, parking sensors, EV chargers, and weather sensors describe the city state. Once they are linked to mobility traces, dwell time, or service usage, they can help us explain why users avoid certain areas.

Data generated by public-service platforms reflects unique behavior. The focus here is not on mobility, but on service use: number of visits, duration of use, type of request, response time, repeat requests, and their link to a specific urban area. A study that discusses the behavioral data of residents identifies sources such as urban traffic data, data from social media, smart devices, and data from public service platforms [4].

An illustration of the smart-city data pipeline is presented in Figure 3, with main data sources and respective behavioral features summarized in Table 6.

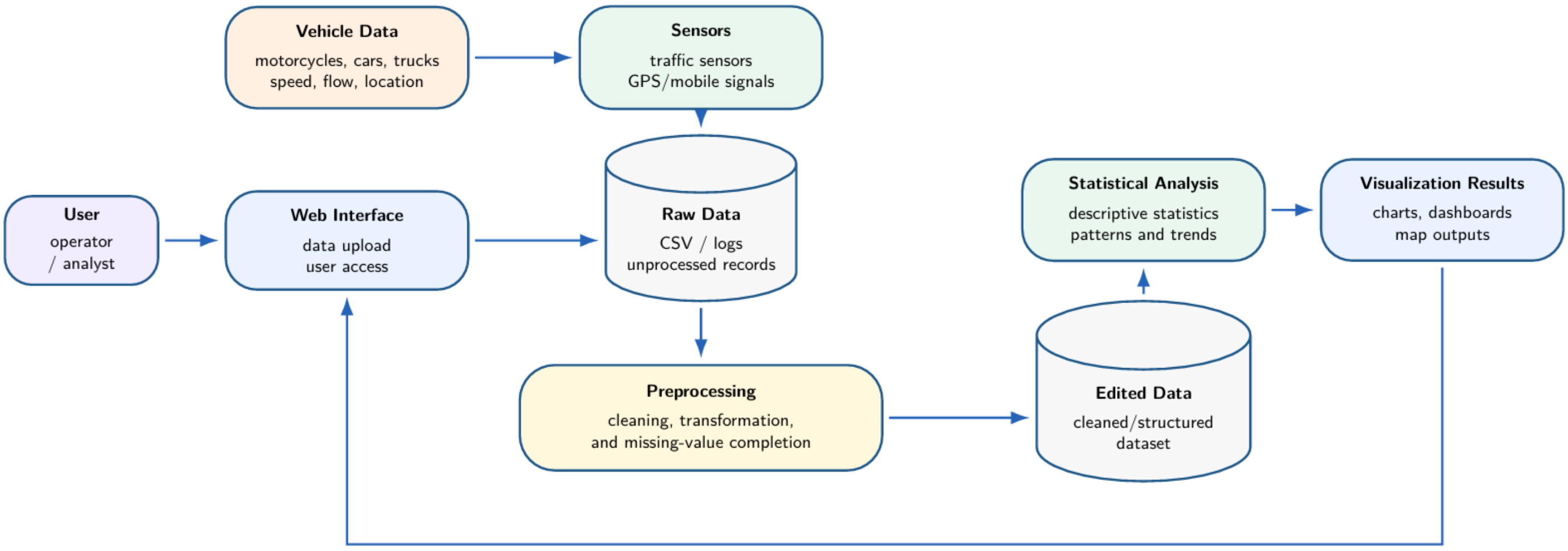

**Figure 3.** Open-access example of a smart-city data pipeline for cleaning, transformation, imputation, statistical analysis, and visualization, similar to [6]

**Table 5.** Conceptual layers of cloud-IoT-edge middleware for smart-city behavioural analytics

| Layer | Main Function | Examples in User Behaviour Analysis |
|---|---|---|
| **Sensing layer** | Collects raw data from physical urban environments and connected devices. | GPS traces, CCTV streams, PoI sensors, parking sensors, EV chargers. |
| **Edge layer** | Performs local, low-latency processing close to the data source. | Anonymisation, object detection, event triggering, local filtering. |
| **Fog layer** | Supports local grouping, coordination, and intermediate processing between edge devices and the cloud. | Neighbourhood-level aggregation, crowd alerts, local mobility summaries. |
| **Cloud layer** | Provides scalable storage, large-scale analysis, and long-term model development. | Historical modelling, model training, city-wide prediction, large-scale behavioural trends. |
| **Middleware layer** | Unifies services, APIs, communication protocols, and data flows across heterogeneous systems. | Service registry, message broker, interoperability mechanisms, service composition. |
| **Context layer** | Adds semantic meaning to raw data by linking it with time, place, activity, and urban context. | Time of day, location, PoI type, activity context, environmental conditions. |
| **Analytics / XAI layer** | Produces behavioural predictions and provides explanations for model outputs. | Mobility prediction, dwell-time classification, anomaly detection, feature importance. |
| **Governance layer** | Manages access control, privacy, accountability, and responsible data use. | Consent management, audit logs, retention policies, privacy rules. |
| **Application layer** | Delivers feedback, decisions, and insights to users, planners, and stakeholders. | Dashboards, AR guidance, planning tools, public-service recommendations. |

**Table 6.** Urban data sources, behavioural features, applications, and associated risks

| **Data Source** | **Data Produced** | **Behavioural Features** | **Applications** | **Risks & Limitations** |
|---|---|---|---|---|
| **GPS / mobile traces** | Location, time, speed | Trajectory, route choice, stop detection | Mobility analysis, tourism, urban planning | Privacy concerns, re-identification risk |
| **Smart cards** | Boarding/alighting data, time, transport line | Commuting patterns, demand clusters | Transport planning, service optimisation | Incomplete picture of wider activity patterns |
| **Bluetooth / Wi-Fi** | Presence, repeat appearance, route traces | Dwell time, pedestrian flow | Public-space analysis, retail, events | Device ambiguity, tracking risk |
| **CCTV / computer vision** | Images, motion data, occupancy | Crowd density, activity zones, abnormal behaviour | Parks, safety, public-space design | Surveillance concerns, bias, occlusion |
| **Social media / LBSN** | Check-ins, posts, likes, geotags | Place preference, visitor flows | Tourism, urban vitality analysis | Sampling bias, self-reporting bias |
| **Public-service platforms** | Visits, requests, duration, service type | Service usage, demand recurrence | E-government, municipal services | Administra tive bias |
| **Environmental sensors** | $NO_2$, $O_3$, PM2.5, noise, weather | Exposure-linked behaviour | Health, mobility, environmental planning | Indirect behavioural inference |
| **Parking / EV sensors** | Occupancy, charging events, duration | Parking demand, charging behaviour | Energy, transport, infrastructure planning | Limited user context |
| **Mobile crowdsensing** | Sensor data, reports, feedback | Citizen-contributed events | Air/noise monitoring, urban maintenance | Consent issues, variable data quality |
| **AR/wearables** | Interaction events, gaze, navigation choices | Engagement, preference, adaptive response | AR city applications, digital twins | High sensitivity of inferred intent |

## 5. Application Domains of User Behaviour Analysis in Smart Cities

When connected to a specific application area, analyzing user behavior in smart cities matters. Behavior is not a mere data category; it consists of things we do all the time in cities: the act of moving somewhere, of staying somewhere, of occupying a public space, consuming energy, using services, participating in public processes, visiting as a tourist, and behaving in the face of risk. The literature on smart cities recognizes that these applications cover transport, energy, environment, health, governance, safety, education, economy, and urban planning [39]. The discourse around smart sustainable cities emphasizes the role of technology in connecting urban intelligence, sustainability, and decision-making [40].

Mobility is the most developed application area. Information related to movement and transport can be collected using information from GPS traces, smart cards, traffic sensors, mobile network data, public transport logs, bike-sharing systems, parking sensors, and navigation apps. In this area, user behavior manifests itself as route choice, departure time, travel frequency, dwell time at terminals, mode choice, and reaction to congestion. When smart-card data is analyzed, commuters' patterns can provide public transport data on routes and the home-work spatial relationship [29], [30].

The second domain refers to conduct at the park, square, museum, shops, and tourism point or area. In this case, spatial dwell time, use of infrastructure, social interaction, PoI preference, and variation by time of day or user type measure behavior. According to a study conducted using CCTV, the design of a place shapes how that place is generally used [5]. In the case of tourism, data comprises check-in, geotagged photographs, application usage, routes, and time spent in a monument [41,42].

The third domain concerns the relationship between energy, behaviour and environmental management. Information on air quality, noise, temperature, traffic emissions, parking occupancy and EV charging becomes behaviourally meaningful when connected to human presence and use. Users' charging behavior means which station he or she chooses, when he or she starts charging, how long he or she connects, the amount of energy drawn, the price sensitivity, and generally what each person prefers, i.e., fast or slow charging [43].

The fourth domain covers public safety, crowd management, and emergency response. Behavioral analysis can support crowd-density estimation, anomaly detection, evacuation routing, emergency alerts, and incident detection. The crowd-modelling literature has highlighted the importance of data-driven methods for understanding collective movement patterns [44], while research on real-time crowd anomaly detection looks at deep learning approaches for safe video surveillance [45].

The fifth domain covers the use of municipal services, citizen feedback, and participatory governance. Here, user behavior shows up as interaction with institutions: submitting requests, using e-government portals, taking part in consultation platforms, reporting problems, and contributing to crowdsensing. Another interesting framework proposes ways citizen voice can be built into smart-city planning [46], while another study [47] looks at forms of citizen participation and engagement across European smart cities [47].

The application domains and their comparative features are summarized in Table 7.

**Table 7.** Application domains and practical examples of user behaviour analysis in smart-city environments

| Domain | Data Sources | Practical Example | Analytics | Sample Metrics |
|---|---|---|---|---|
| **Urban mobility** | GPS traces, smart cards, traffic sensors | Waiting time at stops, route choice, passenger movement patterns | Clustering, prediction, flow analysis | Waiting time, travel time, passenger flow, RMSE |
| **Public spaces/parks** | CCTV, Bluetooth, Wi-Fi, environmental | Dwell time at playgrounds, use of paths, seating-area activity | Computer vision, spatial analytics | Dwell time, occupancy, density, detection latency |
| | or occupancy sensors | | | |
| **Energy/environment** | EV chargers, parking sensors, air/noise sensors | Charging choice, parking behaviour, reaction to pollution exposure | Demand prediction, correlation analysis | Charger occupancy, energy delivered, exposure index |
| **Public safety** | CCTV, crowd sensors, mobile alerts | Crowd congestion, abnormal movement, emergency-related behaviour | Anomaly detection, crowd modelling | F1-score, false alarms, detection latency |
| **Governance/services** | Service logs, feedback apps, participatory maps | Submitting requests, service abandonment, citizen feedback patterns | Sentiment analysis, service analytics | Response time, completion rate, engagement diversity |

## 6. AI and XAI-Driven Behavioural Analytics

User behavior analysis requires more than just data. Data is only valuable when it provides patterns, predictions, and decisions. The use of machine learning, deep learning, spatio-temporal modeling, computer vision, and explainable AI is what behavioural analytics entails. In this case, AI is not an algorithm in isolation. It is a service that lives within cloud-IoT edge middleware.

To analyze behavior, four kinds of questions need to be answered: classification, prediction, anomaly detection, and explanation. Classic machine learning still holds a major role. The classification, segmentation, and prediction tasks make use of SVM, Random Forest, k-means, decision trees, and probabilistic models [4]. Clustering techniques can classify users as commuters, visitors, and high-frequency transit users [27].

Deep learning becomes more important once data is multimodal, time-dependent, or visually complex. Research on Human Activity Recognition shows that CNNs, RNNs, LSTMs, autoencoders, and hybrid architectures have been widely used to recognize activities from sensors and wearables [48,49]. The computer vision literature connects to pedestrian counting, crowd-density estimation, public-space usage analysis, and abnormal-behavior detection [50].

Spatio-temporal modeling research is especially relevant to smart mobility and urban flow prediction. Spatio-Temporal Graph Convolutional Networks treat traffic prediction as a graph-based time-series task [51]. Graph WaveNet extends this with an adaptive dependency matrix, allowing it to learn non-obvious spatiotemporal correlations [52]. These models work well as cloud analytics services for large-scale prediction.

XAI matters here because behavioral analytics shapes decisions in public space. The XAI literature distinguishes between techniques that explain black-box models either locally or globally [53,54]. LIME explains a specific prediction using a simple, interpretable model built around the case being examined [55]. SHAP/TreeSHAP provides a unified approach to feature attribution [56]. With deep networks, assessment of heatmaps and relevance-propagation methods would require some kind of systematic evaluation and not just visual inspection [57-60].

In smart cities, the utilisation of XAI is not just a technical preference but an accountability issue. If a system is able to predict crowd risk, the operator must know whether the prediction was based on the occupancy, time-of-the-day, event schedule, weather, disturbances, and uncertainty in sensors. When a model argues to divert pedestrians or tourists, it needs to state the reason that triggered it. Without explanations, analytics risk turning into a black box to sort and classify people in public space [57-59].

Table 8 maps AI/XAI methods to data type and metric.

**Table 8.** AI and XAI methods for smart-city user behaviour analytics

| **Method** | **Data Type** | **Application** | **Typical Metrics** | **Suitability within Middleware** |
|---|---|---|---|---|
| **k-means / clustering** | Trajectories, smart-card data, service logs | User segmentation, mobility profiles, behavioural grouping | Silhouette score, cluster cohesion | Edge or cloud layer |
| **SVM / Random Forest** | Structured behavioural features | Classification, service-use prediction, behavioural pattern recognition | Accuracy, precision, recall, F1-score | Suitable for early prototypes and lightweight decision modules |
| **LSTM / RNN** | Sequences, temporal usage logs | Next-location prediction, demand forecasting, temporal behaviour modelling | RMSE, MAE, F1-score | Mainly fog or cloud layer |
| **CNN / computer vision** | CCTV streams, image/video frames | Action recognition, occupancy estimation, activity-zone detection | Detection accuracy, mAP, latency | Edge inference, with hardware and latency limitations |
| **Graph WaveNet** | Traffic sensors, spatial graphs, | Traffic forecasting, crowd-flow prediction, spatiotemporal urban modelling | MAE, RMSE, MAPE | Cloud-based analytics service |
| | mobility networks | | | |
| **Anomaly detection** | Crowd data, mobility traces, service logs | Public safety, abnormal demand detection, unusual movement patterns | False-positive rate, recall, detection latency | Edge-cloud hybrid deployment |
| **LIME** | Any classifier | Local explanation of individual model decisions | Fidelity, comprehensibility | XAI service for local explanations |
| **SHAP / TreeSHAP** | Tree models, tabular behavioural data | Feature attribution and explanation of prediction factors | Consistency, stability | XAI audit layer and dashboards |
| **Heatmaps/relevance methods** | Image/video data, | Visual explanation of computer-vision decisions | Perturbation score, explanation plausibility | Vision analytics and visual XAI layer |
| | deep-learning outputs | | | |

## 7. Case Studies and Benchmarking Metrics

The comparison of cloud-IoT middleware solutions to assess user behavior faces a fundamental issue: the studies available do not assess the same thing. Some assess smart-city middleware platforms, others cloud-based IoT platforms, others stream-processing engines, and others domain-specific behavioral analytics. That means comparison can't rely on a single metric - what's needed is a multi-layered benchmarking framework.

InterSCity links smart-city middleware to microservices, scalability, and large-scale simulation [61]. The FIWARE performance evaluation looks at a cloud-based IoT platform for smart cities, examining both vertical and horizontal scalability [62]. SmartSantander operates as a large-scale testbed in a real city [25]. Other useful applications showcase how a big-data platform can manage city data and analytics [62-64]. CityBench and RIoTBench matter because they bring a benchmark-oriented approach to real-time smart-city streams and distributed stream processing [63,65].

The analysis of the behavior of different people, regardless of the purpose, is specifically studied in CCTV camera analysis in urban parks [5], a machine learning-based resident behavior analysis for urban planning [4], an EV-charging-behavior simulator [43], smart tourism through analysis of their digital footprint [42], and citizen participation [46,47]. As per these studies, metrics vary by application domain. For parks, length of stay and density are primary. In EV charging, station occupancy and charging duration are most important. For public services, what matters is completion rate and response time.

The twelve case studies and their evaluation metrics are shown in Table 9, whereas the comparative framework of evaluation metrics is shown in Table 10. The main takeaway from this section is a negative one: there is no single metric capable of adequate evaluation of each cloud–IoT smart-city solution for user behavior analysis. Some metrics tend to come up quite often. At the cloud/API layer, typical parameters are response time, latency, throughput, and requests/events per second. At the hardware level, inference time, CPU/RAM/GPU consumption, and energy consumption matter most. The core measures at the stream-processing level are message rate, query latency, and processing throughput. Analyzing user behavior affects the metrics based on the tasks. Similarly, other important apps are Theodolite [66] in terms of scalability benchmarking of distributed stream processing and in terms of evaluating distributed processing frameworks for IoT smart cities [67].

**Table 9.** Twelve case studies and benchmarking metrics

| No. | Case Study / Benchmark | Domain | Data / Workload | Architecture | Analytics / Processing | Metrics / Benchmarks |
|---|---|---|---|---|---|---|
| 1 | InterSCity scalable platform [61] | Smart-city middleware /parking | Simulated smart-city workload | Microservices, cloud platform | Platform services | Scalability, response under workload |
| 2 | FIWARE performance evaluation [62] | Cloud IoT platform | CoAP/MQTT IoT data | Cloud-based IoT platform | Data ingestion, context management | Latency, throughput, scalability |
| 3 | SmartSantander [25] | City-scale IoT testbed | Urban IoT devices | Large-scale testbed | Experimentation facility | Deployment scale, federation |
| 4 | CiDAP [64] | Big-data analytics | SmartSantander city data | Big-data platform | City data analytics | Data volume, processing capability |
| 5 | CityBench [63] | RDF stream processing benchmark | Smart-city RDF streams | Benchmark suite | Continuous queries | Query latency, throughput |
| 6 | RIoTBench [65] | IoT stream benchmark | 500–10,000 msg/sec workloads | Distributed stream processing | IoT micro-benchmarks | Throughput, latency, scalability |
| 7 | DSP frameworks [67] | Real-time IoT analytics | IoT streams | Distributed stream-processing frameworks | Stream processing | Latency, throughput, resource use |
| 8 | Theodolite [66] | Microservice stream benchmarking | Workload dimensions | Microservices, stream engines | Benchmark execution | Scalability, resource demand |
| 9 | Urban park CCTV [5] | Public-space analytics | CCTV / video data | City cameras and analytics | Computer vision | Occupancy, spatiotemporal activity patterns |
| 10 | Resident behaviour ML [4] | Urban planning / behavioural analytics | Traffic, social, device, and service data | ML-based framework | SVM, RF, LSTM | Prediction performance |
| 11 | EV charging simulator [43] | Energy/mobility | Charging sessions, price signals | Smart-city simulator | Behaviour simulation | Charging duration, station use, price response |
| 12 | Smart tourism imprints [42] | Tourism / POIs | Geotagged traces | Smart-city analytics | PoI identification | PoI frequency, hotspots |

**Table 10.** Comparative framework of evaluation metrics for smart-city middleware and behavioural analytics

| Category | Metrics | Suitable For | Note |
|---|---|---|---|
| **Cloud / API** | Latency, response time, RPS, throughput | FIWARE, dashboards, service APIs | The most common technical metric family. |
| **Edge/hardware** | CPU, RAM, GPU, energy consumption, inference time | CCTV, AR, edge detection | Critical for real-time and resource-constrained use cases. |
| **Network / QoS** | Bandwidth, packet loss, jitter, availability | IoT sensing, MQTT/CoAP communication | Essential in distributed smart-city deployments. |
| **Scalability** | Number of devices, services, users, event rate | InterSCity, Theodolite, FIWARE | Requires controlled workload design. |
| **Stream processing** | Messages per second, query latency, window latency | CityBench, RIoTBench | Useful for continuous behavioural events and real-time data streams. |
| **Data quality** | Missing rate, imputation error, freshness | Dashboards, sensor fusion, urban data integration | Often underrated in smart-city evaluation studies. |
| **ML performance** | Accuracy, precision, recall, F1-score, RMSE | Classification and prediction tasks | Not sufficient without domain-specific evaluation. |
| **Behavioural metrics** | Dwell time, occupancy, PoI visits, crowd density | Parks, tourism, mobility, public-space analysis | Domain-specific and difficult to compare across studies. |
| **XAI / governance** | Fidelity, stability, auditability, consent compliance | High-risk smart-city AI and explainable analytics | Still poorly standardised in the literature. |

## 8. Ethical, Privacy and Governance Challenges

The real-time city relies on a massive computational assembly of urban space, where sensors, platforms and data infrastructures generate a constant flow of data [68]. The movement towards viewing the city from a data system perspective is not neutral. It shapes how governments envision space, citizenship, and urban priorities. When the city is simply seen as an optimization problem, user behaviour gets treated as something to predict and control.

The primary concern is invasion of privacy. Smart-city datasets such as GPS traces, Wi-Fi/Bluetooth detections, smart-card histories, public-service logs, and features derived from CCTV and AR interaction events can provide indications about the routines, sensitive visits, social links, preferences, and vulnerabilities [7,8]. At the level of IoT, privacy is a function of security, as confidentiality, integrity, authentication, access control, trust, and user privacy should be viewed as one single integrated problem [69,70].

Anonymization is important, but it is not sufficient. The principle of k-anonymity states that a record should be indistinguishable from at least $k - 1$ other records concerning quasi-identifiers [71]. Smart city data, on the other hand, is spatiotemporal and repetitive; thus, distinguishing can occur from these unique patterns. Differential privacy provides a stronger mathematical formulation that limits the risk of any one individual's presence in a dataset [72].

The second issue pertains to consent. Smart city consent is not like the one needed to use a mobile app. A person could be in a situation in a public space where there are cameras, sensors, or Wi-Fi counters in place, with no opting out. The standard notice-and-consent model is usually insufficient in this situation. What is required in its place is contextual consent, layered notifications, registers of sensors, and collective accountability mechanisms.

There is also an issue of transparency in the algorithms. Producing predictions about human behavior is not sufficient to be technically accurate; it must also be explainable and auditable [73,74]. General principles of AI ethics point to transparency, justice/fairness, non-maleficence, accountability, and privacy [75,76]. The operational requirements that arise from these principles in smart-city middleware should include: explanation logs, bias checks, model cards, dataset documentation, role-based access, privacy impact assessments, and citizen-facing explanations [77,78].

The fourth problem is democratic involvement. There are many citizen-oriented smart-city projects that are assigning citizens to the role of user, consumer, tester, or even data provider instead of treating them as rights-holders and co-designers of policy [79,80]. Participation cannot simply mean citizens "giving" data. They should participate in deciding on which data should be collected, for what purpose, at what risk, and for what benefit.

## 9. Proposed Framework: XAI-EdgeSOA Middleware for Smart-City Behaviour Analysis

The synthesis of literature shows that an analysis of user behaviour in smart cities must not depend on isolated applications or closed verticals. The sources of data are heterogeneous, the processing must happen over heterogeneous levels of computational proximity, applications are either real-time or near-real-time, and model decisions need to be interpretable, accountable, and auditable. This prompted the draft of the XAI-EdgeSOA Middleware Framework, a.k.a. X-EdgeCity Middleware.

The proposed framework integrates IoT sensing, edge fog/cloud computing, SOA, AI/XAI enabled behavior analytics, and privacy governance. The main position is that a smart city does not simply require more data, but rather it requires controlled middleware that determines where data are collected, where they are processed, which service consumes this data, how the algorithmic output is explained, and which evaluation metric is logged for which decision.

The selection of SOA/microservices is highly impactful in this case. Microservices are beneficial because they are scalable, maintainable, can be deployed independently, and allow for independent technology choice. However, they complicate distributed communication, monitoring, testing, and operations [81,82]. The city needs modularity without runaway operational complexity for smart-city middleware.

The proposed X-EdgeCity Middleware architecture is shown in Figure 4. In particular, the suggested framework is organized into 7 layers:

1. Sensing Layer
2. Edge Preprocessing Layer
3. Event/Data Bus
4. SOA/Microservices
5. Cloud Analytics
6. XAI & Governance
7. Application/Feedback Layer.

In traditional cloud applications, unprocessed data is sent to the cloud for centralized storage and analysis. This causes latency, bandwidth, privacy, and operational risks, which is a major issue for smart-city behavioral applications. According to the Edge Intelligence literature, moving AI functions to the edge reduces latency and enables local usage of data [83,84].

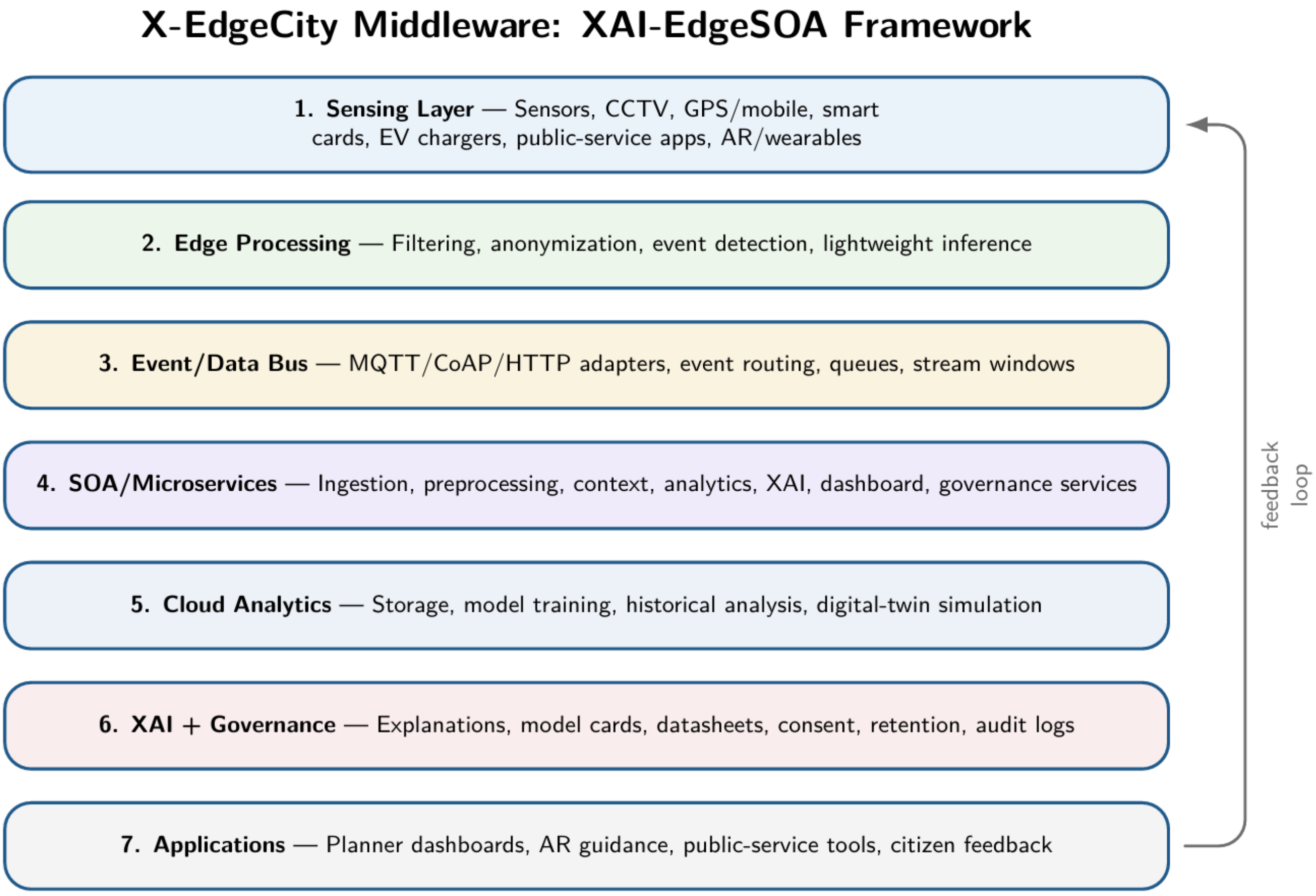

**Figure 4.** Proposed X-EdgeCity Middleware Framework

The design allows for a phased transition from monolithic to microservices. A single code base defines many smart-city applications. As such, moving to cloud-native and microservice architectures can support scalability and DevOps, as long as the breakup is done in a controlled way [85].

The XAI Layer acts as an independent service. Every prediction classified as medium-risk or high-risk needs to come with a model identifier, dataset version, input feature summary, confidence score, explanation method, top contributing features, uncertainty indicator, and policy status. Tabular models can use SHAP-like explanations [56]. Local explanations can use LIME-like methods [55]. Datasheets for Datasets and Model Cards can be built in as mandatory metadata objects [86,87].

## 10. Discussion

This review shows that analyzing user behavior in smart cities can't be treated purely as a technical issue, nor as a simple case of applying machine learning to urban data. The convergence of spatial engineering, computational intelligence, and ethical governance will facilitate a new era of infrastructure development and spatial data analysis. The key finding is that the smart city does not just produce data; it produces signs, cues, and traces of interpretation. To borrow an analogy from linguistics: raw sensor events are like a form of urban syntax, a sequence of signals that do not bear automatic meaning. The meaning of events only emerges when they connect with place, time, purpose, context, social use, and institutional framing. The framework integrates IoT sensing, edge/fog/cloud computing, service-oriented architecture, AI/XAI driven behavioral analytics, and privacy-aware governance. The central point is that a smart city needs more than just data; it needs a middleware entity to orchestrate and coordinate the processes to determine where the data is being collected, where it is processed, which service consumes it, how the algorithmic output explanation takes place, and which evaluation metric is logged for which decision.

Findings reveal that closed, monolithic, vertical information systems cannot be the one-stop smart-city solution at the architectural level. Studies of InterSCity, FIWARE, SmartSantander, and microservice-based platforms indicate that interoperability, extensibility, and orchestration ability of services are some of the most basic requirements for any serious urban infrastructure [1,25,61,62]. The SOA/microservices solution that will be developed as part of X-EdgeCity Middleware can provide just this decomposition: ingestion, preprocessing, context enrichment, analytics, XAI, visualization, governance [81,82,85].

The case studies and benchmarks highlighted a genuine methodological gap: there is currently no common evaluation metric in the literature for studying smart-city behaviour. At the platform level, latency, throughput, scalability, requests/events per second, and resource utilization dominate [61-75]. For AI models, accuracy, precision, recall, F1, RMSE, or MAE are the norm [48-52]. In the behavioral domains, what shows up is dwell time, occupancy, crowd density, PoI visits, charging duration, waiting time, or citizen engagement [5], [42-47]. This lack of a shared metric shouldn't be seen as a weakness to be fixed with some single super-metric. A more realistic answer is a multi-metric evaluation model.

The question of ethical governance is central. The most dangerous form of surveillance isn't always the visible camera. It's the quiet algorithmic sorting: which area gets flagged as problematic, which user gets labeled as deviant, which behavior gets marked as abnormal, which route gets promoted and which gets hidden [68-80]. In linguistic terms, the problem is the power of naming: the system doesn't just describe the city; it categorizes it. And whoever categorizes intervenes.

The X-EdgeCity Middleware proposal addresses this in three ways. First, it limits data collection through edge preprocessing and data minimization. Second, it makes decisions auditable through XAI objects, model cards, datasheets, and audit logs [86,87]. Third, it builds governance in as a functional layer rather than treating it as an external policy statement bolted on afterward. Ethical principles need to be translated into policies, access rules, retention timers, explanation requirements, and benchmark records.

The primary theoretical conclusion states that a smart city is an interpretive system in itself. The city gathers signs, but the systems determine the meaning of the signs. The engineering side manages the protocols, APIs, models, and metrics. Data does not interpret itself; rather, 'someone' interprets it. In terms of social impact, we can ask ourselves who benefits from that reading, and who suffers from it.

## 11. Conclusions

This particular review analysed the question of user behaviour in smart cities from cloud, IoT edge middleware, machine learning, explainability, and ethical governance perspectives. The primary takeaway is that one cannot understand a smart city as the straightforward sum of sensors, dashboards, and algorithms. It is a socio-technical system with many layers where things like the physical environment, digital traces, computing infrastructure, analytical models, and institutional decisions come together in one powerful location to produce new urban knowledge.

The first research question considered cloud–IoT, edge/fog, and middleware architectures that can enable user behavior analysis. The review found multi-layered distributed service-oriented architectures to be the most suitable. The cloud is a major enabler in terms of its capabilities in scale and computing power, either for historical analytics or training. Moreover, edge and fog computing work alongside the cloud to facilitate local processing and faster action when sending and receiving messages. They also lower bandwidth requirements by limiting the transfer of raw data [18-20,83,84].

The second research question concerned evaluation metrics. Looking at the case studies of this study and related benchmarking papers, it is evident that there's no single universal metric for evaluating smart-city user behavior analysis. At the infrastructure and middleware level, the metrics that keep coming up are latency, response time, throughput, requests/events per second, scalability, resource utilization, and availability. At the machine-learning level, the most common metrics are accuracy, precision, recall, F1-score, AUC, MAE, MSE, and RMSE [48-52]. At the application level, metrics are domain-specific: dwell time and occupancy, waiting time and passenger flow, charging duration and station occupancy, crowd density and false-alarm rate, completion rate and engagement diversity.

The novel middleware framework design was the focus of the third research question. The contribution proposed is the XAI-EdgeSOA Middleware Framework (X-Edge-City Middleware). This system takes advantage of various functionalities integrated into the solution, including IoT sensing and edge preprocessing to fog coordination and more. The structure of its architecture relies on the premise that behavioural analytics must be executed at the right level of computation. Moreover, the ethical dimension is the most crucial conclusion. An intelligent city may aid in service enhancement but can equally be an instrument of urban surveillance. It is not just the technology that makes a difference, but also the rules created about its use. Functional requirements of the system itself, not extras bolted on from without, include privacy by design, data minimization, purpose limitation, retention control, access governance, bias auditing and participatory oversight [7,8,68-80].

When linguistics is brought together with engineering, we get a useful closing way of thinking about all this: the smart city is a machine for the production of urban signs. Sensors generate signals, middleware organizes signals into structures, models transform them into interpretations, and institutions turn interpretations into decisions. As such, a smart city's quality depends on the quality of this interpretive chain. This means that the final endpoint of a cloud-IoT edge middleware is not only in the speed of data processing but in whether it can generate meaningful interpretations without undermining the rights, trust, and democratic oversight.

**Supplementary Materials:** Not applicable.

**Author Contributions:** All authors contributed equally to all stages of the manuscript.

**Funding:** This research received no external funding.

**Data Availability Statement:** Not applicable.

**Acknowledgments:** The authors have reviewed and edited the output and take full responsibility for the content of this publication.

**Conflicts of Interest:** The authors declare no conflicts of interest

## References

1. Batista DM, Goldman A, Hirata R, Kon F, Costa FM, Endler M. Interscity: Addressing future internet research challenges for smart cities. In 2016 7th International Conference on the Network of the Future (NOF), 2016 Nov 16 (pp. 1-6). IEEE. https://doi.org/10.1109/NOF.2016.7810114
2. Goumopoulos C. Smart city middleware: A survey and a conceptual framework. IEEE Access. 2024 Jan 3;12:4015-47. https://doi.org/10.1109/ACCESS.2023.3349376
3. Luo Z, Su J, Wu W, Dai X. UAV-enabled mobile edge computing: a comprehensive survey on architectures, optimization techniques, and multi-scenario applications. Cluster Computing. 2026 Aug;29(10):621. https://doi.org/10.1007/s10586-026-06455-5
4. Peng Z, Zhao Y, Luo Z. Machine Learning-Based Analysis of Resident Behaviour Patterns and Urban Planning in Smart Cities. In2024 3rd International Conference on Smart City Challenges & Outcomes for Urban Transformation (SCOUT), 2024 Jul 26 (pp. 52-57). IEEE. https://doi.org/10.1109/SCOUT64349.2024.00021
5. Gravitz-Sela S, Shach-Pinsly D, Bryt O, Plaut P. Leveraging City Cameras for Human Behavior Analysis in Urban Parks: A Smart City Perspective. Sustainability. 2025 Jan 22;17(3):865. https://doi.org/10.3390/su17030865
6. Karampakakis P, Ioakeimidou D, Chatzimisios P, Tsintotas KA. A web-based application for smart city data analysis and visualization. Future Internet. 2025 May 13;17(5):217. https://doi.org/10.3390/fi17050217
7. Emami-Naeini P, Breda J, Dai W, Kohno T, Laine K, Patel S, Roesner F. Understanding people's concerns and attitudes toward smart cities. In Proceedings of the 2023 CHI Conference on Human Factors in Computing Systems, 2023 Apr 19 (pp. 1-24). https://doi.org/10.1145/3544548.3581558
8. Van Zoonen L. Privacy concerns in smart cities. Government Information Quarterly. 2016 Jul 1;33(3):472-80. https://doi.org/10.1016/j.giq.2016.06.004
9. Page MJ, McKenzie JE, Bossuyt PM, Boutron I, Hoffmann TC, Mulrow CD, Shamseer L, Tetzlaff JM, Akl EA, Brennan SE, Chou R. The PRISMA 2020 statement: an updated guideline for reporting systematic reviews. bmj. 2021 Mar 29;372. https://doi.org/10.1136/bmj.n71
10. Tricco AC, Lillie E, Zarin W, O'Brien KK, Colquhoun H, Levac D, Moher D, Peters MD, Horsley T, Weeks L, Hempel S. PRISMA extension for scoping reviews (PRISMA-ScR): checklist and explanation. Annals of Internal Medicine. 2018 Oct 2;169(7):467-73. https://doi.org/10.7326/M18-0850
11. Arksey H, O'Malley L. Scoping studies: towards a methodological framework. International journal of social research methodology. 2005 Feb 1;8(1):19-32. https://doi.org/10.1080/1364557032000119616
12. Levac D, Colquhoun H, O'Brien KK. Scoping studies: advancing the methodology. Implementation Science. 2010 Sep 20;5(1):69. https://doi.org/10.1186/1748-5908-5-69
13. Zanella A, Bui N, Castellani A, Vangelista L, Zorzi M. Internet of things for smart cities. IEEE Internet of Things Journal. 2014 Feb 14;1(1):22-32. https://doi.org/10.1109/JIOT.2014.2306328
14. Pittri H, Godawatte GA, Forster AM. Internet of Things in construction supply chain management: a review of applications, challenges, integration pathways and future directions. Journal of Engineering, Design and Technology. 2026 Jan 29:1-26. https://doi.org/10.1108/JEDT-09-2025-0487
15. Razzaque MA, Milojevic-Jevric M, Palade A, Clarke S. Middleware for Internet of Things: a survey. IEEE Internet of Things Journal. 2015 Nov 9;3(1):70-95. https://doi.org/10.1109/JIOT.2015.2498900
16. Botta A, De Donato W, Persico V, Pescapé A. Integration of cloud computing and Internet of Things: a survey. Future Generation Computer Systems. 2016 Mar 1;56:684-700. https://doi.org/10.1016/j.future.2015.09.021
17. Bonomi F, Milito R, Zhu J, Addepalli S. Fog computing and its role in the Internet of things. In Proceedings of the first edition of the MCC workshop on Mobile cloud computing, 2012 Aug 17 (pp. 13-16). https://doi.org/10.1145/2342509.2342513
18. Taleb I, Guillaume JL, Duthil B. A survey on services placement algorithms in integrated cloud-fog/edge computing. ACM Computing Surveys. 2025 Jun 11;57(11):1-36. https://doi.org/10.1145/3729214
19. Atwal S, Singh D, Chhabra A. Exploring the potential of artificial intelligence and machine learning in fog computing, cloud computing and edge computing: Benefits and challenges. In2025 International Conference on Intelligent Control, Computing and Communications (IC3), 2025 Feb 13 (pp. 199-207). IEEE. https://doi.org/10.1109/IC363308.2025.10957326
20. Ahmed ZR, Askar S, Hussein DH, Ibrahim MA. Fog computing challenges and opportunities in IoT networks: A review. Procedia Computer Science. 2025 Jan 1;259:1749-64. https://doi.org/10.1016/j.procs.2025.04.130
21. Vishwakarma AK, Chaurasia S, Kumar K, Singh YN, Chaurasia R. Internet of things technology, research, and challenges: a survey. Multimedia Tools and Applications. 2025 Mar;84(11):8455-90. https://doi.org/10.1007/s11042-024-19278-6
22. Kufakunesu R, Myburgh H, De Freitas A. The Internet of battle things: a survey on communication challenges and recent solutions. Discover Internet of Things. 2025 Jan 10;5(1):3. https://doi.org/10.1007/s43926-025-00093-w

23. Kumara I, Kaplan H, Owotogbe J, Tamburri DA, van den Heuvel WJ. Large language models for service-oriented computing (llm4soc): review and research directions. In Symposium and Summer School on Service-Oriented Computing 2025, Jun 16 (pp. 90-113). Cham: Springer Nature Switzerland. https://doi.org/10.1007/978-3-032-07313-6_6
24. Singh RP, Thakur M, Razavi SM, Sankhla S, Singh KK, Limbasiya IH. Monolithic and Microservice Architecture: A Sustainable Approach. In2025 3rd IEEE International Conference on Industrial Electronics: Developments & Applications (ICIDeA) 2025 Feb 21 (pp. 1-6). IEEE. https://doi.org/10.1109/ICIDeA64800.2025.10962984
25. Sanchez L, Muñoz L, Galache JA, Sotres P, Santana JR, Gutierrez V, Ramdhany R, Gluhak A, Krco S, Theodoridis E, Pfisterer D. SmartSantander: IoT experimentation over a smart city testbed. Computer Networks. 2014 Mar 14;61:217-38. https://doi.org/10.1016/j.bjp.2013.12.020
26. Li Z, Xia L, Ren X, Tang J, Chen T, Xu Y, Huang C. Urban computing in the era of large language models. ACM Transactions on Intelligent Systems and Technology. 2025 Nov 24;16(6):1-43. https://doi.org/10.1145/3768163
27. Aremu DR. Machine Learning and Data Mining: An Overview for Emerging Trend in Fourth Industrial Revolution. Robotic Technology, Artificial Intelligence and the 4IR Libraries. 2026 Feb 2:287. https://books.google.gr/books?hl=en&lr=&id=T3aaEQAAQBAJ&oi=fnd&pg=PA287&dq=Data+Mining:+An+Overview&ots=FMzMfBWlpe&sig=v20OBnb-rJlLbzhNOoOKbzqCYR8&redir_esc=y#v=onepage&q&f=false [Access Date: 08/09/2026]
28. Li J, Ranjan R, Wang Y, Yan J, Huang X. A survey on Urban spatial computing system. Computing. 2025 Sep;107(9):184. https://doi.org/10.1007/s00607-025-01536-6
29. Wang Y, Choudhury C, Hancock TO, Wang Y. Investigating the changes in residential location and commute patterns during the pandemic using smart card data. Travel Behaviour and Society. 2025 Apr 1;39:100964. https://doi.org/10.1016/j.tbs.2024.100964
30. Cats O. Identifying human mobility patterns using smart card data. Transport Reviews. 2024 Jan 2;44(1):213-43. https://doi.org/10.1080/01441647.2023.2251688
31. Wang J, Xia M, Xie Y, Shi C. Multimodal sensor fusion-based lightweight pedestrian wearable system for continuous health monitoring and location tracking. IEEE Sensors Journal. 2025 Aug 5. https://doi.org/10.1109/JSEN.2025.3593999
32. Zhang S, He N, Li Z, Fu Q, Liang W, Xiong NN. A Survey on Mobile Crowd Sensing: Concepts, Applications, Technologies, and Future Directions. IEEE Internet of Things Journal. 2025 Oct 31;12(24):52090-109. https://doi.org/10.1109/JIOT.2025.3627513
33. Lloret Á, Peral J, Ferrández A, Auladell M, Muñoz R. A data-driven framework for digital transformation in smart cities: integrating AI, dashboards, and IoT readiness. Sensors. 2025 Aug 20;25(16):5179. https://doi.org/10.3390/s25165179
34. Gutiérrez A, Domènech A, Zaragozí B, Miravet D. Profiling tourists' use of public transport through smart travel card data. Journal of Transport Geography. 2020 Oct 1;88:102820. https://doi.org/10.1016/j.jtrangeo.2020.102820
35. Kerrache CA, Sahraoui Y, Calafate CT, Vegni AM, editors. Mobile Crowdsensing and Remote Sensing in Smart Cities. Springer; 2025. https://link.springer.com/book/10.1007/978-3-031-72732-0 [Access Date: 08/09/2026]
36. Shamroukh M, Natapov A, Larimian T. Reliability and scalability of pedestrian monitoring practices: A systematic review. IEEE Access. 2025 Aug 15. https://doi.org/10.1109/ACCESS.2025.3599283
37. Niu H, Silva EA. Understanding temporal and spatial patterns of urban activities across demographic groups through geotagged social media data. Computers, Environment and Urban Systems. 2023 Mar 1;100:101934. https://doi.org/10.1016/j.compenvurbsys.2022.101934
38. Bernabeu-Bautista Á, Serrano-Estrada L, Perez-Sanchez VR, Martí P. The geography of social media data in urban areas: Representativeness and complementarity. ISPRS International Journal of Geo-Information. 2021 Nov 3;10(11):747. https://doi.org/10.3390/ijgi10110747
39. Trigka M, Dritsas E. Edge and cloud computing in smart cities. Future Internet. 2025 Mar 6;17(3):118. https://doi.org/10.3390/fi17030118
40. De Silva D, Mills N, Moraliyage H, Rathnayaka P, Wishart S, Jennings A. Responsible artificial intelligence hyper-automation with generative AI agents for sustainable cities of the future. Smart Cities. 2025 Feb 17;8(1):34. https://doi.org/10.3390/smartcities8010034
41. Chen S, Chen Y, Zhang K, Law R. Behavioural theories in smart tourism. Current issues in tourism. 2026 Apr 18;29(8):1644-67. https://doi.org/10.1080/13683500.2025.2542968
42. Encalada L, Boavida-Portugal I, Cardoso Ferreira C, Rocha J. Identifying tourist places of interest based on digital imprints: Towards a sustainable smart city. Sustainability. 2017 Dec 13;9(12):2317. https://doi.org/10.3390/su9122317

43. Canizes B, Soares J, Costa A, Pinto T, Lezama F, Novais P, Vale Z. Electric vehicles' user charging behaviour simulator for a smart city. Energies. 2019 Apr 18;12(8):1470. https://doi.org/10.3390/en12081470
44. Hachaj T, Wąs J. An insightful data-driven crowd simulation model based on rough sets. Information Sciences. 2025 Feb 1;692:121670. https://doi.org/10.1016/j.ins.2024.121670
45. Rezaee K, Rezakhani SM, Khosravi MR, Moghimi MK. A survey on deep learning-based real-time crowd anomaly detection for secure distributed video surveillance. Personal and Ubiquitous Computing. 2024 Feb;28(1):135-51. https://doi.org/10.1007/s00779-021-01586-5
46. Landa Oregi I, Urra-Uriarte S, Gonzalez Ochoantesana I, Rodríguez MA, Molina-Costa P. Enhancing citizen participation in citizen-centered smart cities: Insights from two European case studies. Urban Science. 2025 Apr 24;9(5):140. https://doi.org/10.3390/urbansci9050140
47. Kummitha RK. Smart city governance: assessing modes of active citizen engagement. Regional Studies. 2025 Dec 31;59(1):2399262. https://doi.org/10.1080/00343404.2024.2399262
48. Gu F, Chung MH, Chignell M, Valaee S, Zhou B, Liu X. A survey on deep learning for human activity recognition. ACM Computing Surveys (CSUR). 2021 Oct 4;54(8):1-34. https://doi.org/10.1145/3472290
49. Dhekane SG, Ploetz T. Transfer learning in sensor-based human activity recognition: A survey. ACM Computing Surveys. 2025 Mar 23;57(8):1-39. https://doi.org/10.1145/3717608
50. AlShami AK, Rabinowitz R, Lam K, Shleibik Y, Mersha M, Boult T, Kalita J. Smart-vision: survey of modern action recognition techniques in vision. Multimedia tools and applications. 2025 Aug;84(27):32705-76. https://doi.org/10.1007/s11042-024-20484-5
51. He P, Di H, He C, Zhang X, Liu C, Zhou S. Spatio-temporal graph convolutional networks with multi-source fusion for predicting excavation deformation. Advanced Engineering Informatics. 2025 Nov 1;68:103722. https://doi.org/10.1016/j.aei.2025.103722
52. Pan Y, Wang H, Ma M, Wang P. Enhancing Spatial-Temporal Prediction Models with Dynamic Causal Graphs. ACM Transactions on Spatial Algorithms and Systems. 2025 Dec 17;11(4):1-22. https://doi.org/10.1145/3777547
53. Guidotti R, Monreale A, Ruggieri S, Turini F, Giannotti F, Pedreschi D. A survey of methods for explaining black box models. ACM Computing Surveys (CSUR). 2018 Aug 22;51(5):1-42. https://doi.org/10.1145/3236009
54. Somvanshi S, Islam MM, Rafe A, Tusti AG, Chakraborty A, Baitullah A, Chowdhury TI, Alnawmasi N, Dutta A, Das S. Bridging the black box: a survey on mechanistic interpretability in AI. ACM Computing Surveys. 2026 Feb 4;58(8):1-35. https://doi.org/10.1145/3787104
55. Ribeiro MT, Singh S, Guestrin C. " Why should i trust you?" Explaining the predictions of any classifier. In Proceedings of the 22nd ACM SIGKDD International Conference on knowledge discovery and data mining 2016, Aug 13 (pp. 1135-1144). https://doi.org/10.1145/2939672.2939778
56. Lundberg SM, Erion G, Chen H, DeGrave A, Prutkin JM, Nair B, Katz R, Himmelfarb J, Bansal N, Lee SI. From local explanations to global understanding with explainable AI for trees. Nature Machine Intelligence. 2020 Jan;2(1):56-67. https://doi.org/10.1038/s42256-019-0138-9
57. Khan D, Khan N, Ullah S. Harnessing hybrid intelligence and explainable AI for urban growth prediction: A Data-Driven framework for sustainable cities. Environment, development and sustainability. 2025 Sep 20:1-40. https://doi.org/10.1007/s10668-025-06860-7
58. Almaazmi KI, Almheiri SJ, Khan MA, Shah AA, Abbas S, Ahmad M. Enhancing smart city sustainability with explainable federated learning for vehicular energy control. Scientific Reports. 2025 Jul 4;15(1):23888. https://doi.org/10.1038/s41598-025-07844-3
59. Ebrahimi F, Javidan R, Akbari R, Hosseini Y. Intrusion detection in the Internet of things using convolutional neural networks: an explainable AI approach. Cybersecurity. 2025 Sep 5;8(1):66. https://doi.org/10.1186/s42400-025-00369-2
60. Samek W, Binder A, Montavon G, Lapuschkin S, Müller KR. Evaluating the visualization of what a deep neural network has learned. IEEE transactions on neural networks and learning systems. 2016 Aug 25;28(11):2660-73. https://doi.org/10.1109/TNNLS.2016.2599820
61. Del Esposte AD, Santana EF, Kanashiro L, Costa FM, Braghetto KR, Lago N, Kon F. Design and evaluation of a scalable smart city software platform with large-scale simulations. Future Generation Computer Systems. 2019 Apr 1;93:427-41. https://doi.org/10.1016/j.future.2018.10.026
62. Araujo V, Mitra K, Saguna S, Åhlund C. Performance evaluation of FIWARE: A cloud-based IoT platform for smart cities. Journal of Parallel and Distributed Computing. 2019 Oct 1;132:250-61. https://doi.org/10.1016/j.jpdc.2018.12.010

63. Ali MI, Gao F, Mileo A. Citybench: A configurable benchmark to evaluate rsp engines using smart city datasets. In International Semantic Web Conference 2015, Oct 11 (pp. 374-389). Cham: Springer International Publishing. https://doi.org/10.1007/978-3-319-25010-6_25
64. Cheng B, Longo S, Cirillo F, Bauer M, Kovacs E. Building a big data platform for smart cities: Experience and lessons from Santander. In 2015 IEEE International Congress on Big Data, 2015 Jun 27 (pp. 592-599). IEEE. https://doi.org/10.1109/BigDataCongress.2015.91
65. Shukla A, Chaturvedi S, Simmhan Y. Riotbench: An IoT benchmark for distributed stream processing systems. Concurrency and Computation: Practice and Experience. 2017 Nov 10;29(21):e4257. https://doi.org/10.1002/cpe.4257
66. Henning S, Hasselbring W. Theodolite: Scalability benchmarking of distributed stream processing engines in microservice architectures. Big Data Research. 2021 Jul 15;25:100209. https://doi.org/10.1016/j.bdr.2021.100209
67. Nasiri H, Nasehi S, Goudarzi M. Evaluation of distributed stream processing frameworks for IoT applications in Smart Cities. Journal of Big Data. 2019 Jun 11;6(1):52. https://doi.org/10.1186/s40537-019-0215-2
68. Miftah M, Desrianti DI, Septiani N, Fauzi AY, Williams C. Big data analytics for smart cities: Optimizing urban traffic management using real-time data processing. Journal of computer science and technology application. 2025 Feb 21;2(1):14-23. https://doi.org/10.33050/corisinta.v2i1.56
69. Herriger C, Merlo O, Eisingerich AB, Arigayota AR. Context-contingent privacy concerns and exploration of the privacy paradox in the age of AI, augmented reality, big data, and the internet of things: systematic review. Journal of medical Internet research. 2025 May 14;27:e71951. https://doi.org/10.2196/71951
70. Zhao D, Li Q, Zou Q, Xiao J, Wu K, Li R, Lv J, Jiang Y, Li K. Security and privacy in smart homes: Challenges and latest developments. In Advances in the Internet of Things, 2025 Jul 17 (pp. 36-55). CRC Press. https://www.taylorfrancis.com/chapters/edit/10.1201/9781003506638-3/security-privacy-smart-homes-dan-zhao-qing-li-qingsong-zou-jingyu-xiao-kaidong-wu-ruoyu-li-jianhui-lv-yong-jiang-keqin-li [Access Date: 08/09/2026]
71. Sweeney L. k-anonymity: A model for protecting privacy. International journal of uncertainty, fuzziness and knowledge-based systems. 2002 Oct;10(05):557-70. https://doi.org/10.1142/S0218488502001648
72. Dwork C. Differential privacy. In International colloquium on automata, languages, and programming 2006 Jul 10 (pp. 1-12). Berlin, Heidelberg: Springer Berlin Heidelberg. https://doi.org/10.1007/11787006_1
73. Ananny M, Crawford K. Seeing without knowing: Limitations of the transparency ideal and its application to algorithmic accountability. new media & society. 2018 Mar;20(3):973-89. https://doi.org/10.1177/1461444816676645
74. Mittelstadt BD, Allo P, Taddeo M, Wachter S, Floridi L. The ethics of algorithms: Mapping the debate. Big data & society. 2016 Nov;3(2):2053951716679679. https://doi.org/10.1177/2053951716679679
75. Jobin A, Ienca M, Vayena E. The global landscape of AI ethics guidelines. Nature Machine Intelligence. 2019 Sep;1(9):389-99. https://doi.org/10.1038/s42256-019-0088-2
76. Floridi L, Cowls J. A unified framework of five principles for AI in society. Machine learning and the city: Applications in architecture and urban design. 2022 May 21:535-45. https://doi.org/10.1002/9781119815075.ch45
77. Goodman B, Flaxman S. European Union regulations on algorithmic decision-making and a “right to explanation”. AI Magazine. 2017 Oct 2;38(3):50-7. https://doi.org/10.1609/aimag.v38i3.2741
78. Silva H, Garcia H. Ethical challenges in large-scale IoT and smart city deployments. Journal of Ethics in Emerging Technologies & Society p-ISSN 3117-5996 and e-ISSN 3117-6003. 2025 Jun 30;2(2):35-42. https://galaxiauniverse.com/index.php/JEETS/article/view/422 [Access Date: 08/09/2026]
79. Hansen MM, Dahiya B. Traffy Fondue: a smart city citizen engagement. Frontiers in Sustainable Cities. 2025 Apr 2;7:1491621. https://doi.org/10.3389/frsc.2025.1491621
80. Srivastava M, Ghazala S. Future-proofing sustainable urban development: harnessing fuzzy logic for smart cities. Fuzzy Logic in Smart Sustainable Cities. 2025 Jun 4:215-57. http://doi.org/10.1201/9781003636724-9
81. Dragoni N, Giallorenzo S, Lafuente AL, Mazzara M, Montesi F, Mustafin R, Safina L. Microservices: yesterday, today, and tomorrow. Present and ulterior software engineering. 2017 Sep 6:195-216. https://doi.org/10.1007/978-3-319-67425-4_12
82. Soldani J, Tamburri DA, Van Den Heuvel WJ. The pains and gains of microservices: A systematic grey literature review. Journal of Systems and Software. 2018 Dec 1;146:215-32. https://doi.org/10.1016/j.jss.2018.09.082
83. Wang T, Guo J, Zhang B, Yang G, Li D. Deploying AI on edge: Advancement and challenges in edge intelligence. Mathematics. 2025 Jun 4;13(11):1878. https://doi.org/10.3390/math13111878
84. Ficili I, Giacobbe M, Tricomi G, Puliafito A. From sensors to data intelligence: Leveraging IoT, cloud, and edge computing with AI. Sensors. 2025 Mar 12;25(6):1763. https://doi.org/10.3390/s25061763

85. Balalaie A, Heydarnoori A, Jamshidi P. Microservices architecture enables devops: Migration to a cloud-native architecture. IEEE Software. 2016 Mar 18;33(3):42-52. https://doi.org/10.1109/MS.2016.64
86. Gebru T, Morgenstern J, Vecchione B, Vaughan JW, Wallach H, Daume HI, Crawford K. Datasheets for datasets. Communications of the ACM. 2021 Nov 19;64(12):86-92. https://dx.doi.org/10.1145/3458723
87. Mitchell M, Wu S, Zaldivar A, Barnes P, Vasserman L, Hutchinson B, Spitzer E, Raji ID, Gebru T. Model cards for model reporting. In Proceedings of the Conference on Fairness, Accountability, and Transparency 2019 Jan 29 (pp. 220-229). https://doi.org/10.1145/3287560.3287596